\documentclass{emulateapj}

\usepackage{apjfonts}
\usepackage{epsfig}
\usepackage{natbib}

\newlength{\colwidth}

\newcommand{\be}{\begin{equation}}
\newcommand{\ee}{\end{equation}}
\newcommand{\bee}{\begin{eqnarray}}
\newcommand{\eee}{\end{eqnarray}}

\newcommand{\msol}{\hbox{${\rm M}_\odot$}}
\newcommand{\rsol}{\hbox{${{\bf r}}_\odot$}}
\newcommand{\vsol}{\hbox{${{\bf v}}_\odot$}}

\newcommand{\rvir}{\hbox{$r_{\rm 200}$}}
\newcommand{\mvir}{\hbox{$M_{\rm vir}$}}
\newcommand{\Mvir}{\hbox{$M_{\rm vir}$}}

\newcommand{\vmax}{{\rm v}_{\rm max}}

\newcommand{\LCDM}{$\Lambda$CDM }

\newcommand{\MMW}{{\rm M}_{\rm MW}}
\newcommand{\Nsubs}{N_{\rm subs}}

\newcommand{\kms}{km $s^{-1}$}
\newcommand{\Msun}{{\rm M}_\odot}

\def\pr{{\rm Pr}}
\def\pars{\mathbf{x}}

\def\data{\mathbf{d}}
\def\datai{d_i}

\newcommand{\hinv}{h^{-1}}
\newcommand{\mpc}{\rm{Mpc}}
\newcommand{\hmpc}{\hinv\mpc}
\newcommand{\kpc}{\rm{kpc}}

\newcommand{\lsim}{\lower.5ex\hbox{\ltsima}}
\newcommand{\gsim}{\lower.5ex\hbox{\gtsima}}
\newcommand{\ltsima}{$\; \buildrel < \over \sim \;$}
\newcommand{\gtsima}{$\; \buildrel > \over \sim \;$}

\def\MassEstimate{1.2^{+0.7}_{-0.4}{\rm\,(stat.)}^{+0.3}_{-0.3}{\rm\,(sys.)} \times 10^{12}}
\def\logMassEstimate{12.2 \pm 0.1}
\def\cEstimate{11 \pm 2}
\def\cpopEstimate{8.7 \pm 3.5}
\def\accTimeProb{72\%}
\def\accTimeLimit{1~Gyr}
\def\SimultaneousAccFraction{50\%}
\def\Ngoodvolume{$500 \mpc^{3}$}

\usepackage[usenames]{color}

\shortauthors{Busha et al}
\shorttitle{Milky Way Properties from Observations of the Magellanic Clouds}

\begin{document}

\title{The Mass Distribution and Assembly of the Milky Way \\
from the Properties of the Magellanic Clouds}

\author{Michael T. Busha\altaffilmark{1,2}} 
\author{Philip J. Marshall\altaffilmark{1,3}}
\author{Risa H. Wechsler\altaffilmark{1,4}}
\author{Anatoly Klypin\altaffilmark{5}}
\author{Joel Primack\altaffilmark{6}}
\altaffiltext{1}{Kavli Institute for Particle Astrophysics and Cosmology,
Physics Department, Stanford University, Stanford, CA, 94305, USA
{\tt mbusha@physik.uzh.ch, pjm@slac.stanford.edu, rwechsler@stanford.edu}}
\altaffiltext{2}{Institute for Theoretical Physics, University of
  Z\"urich, 8057 Z\"urich, Switzerland}  
\altaffiltext{3}{Department of Physics, Oxford University, Oxford, OX1 3RH, UK.} 
\altaffiltext{4}{Particle and Particle Astrophysics Department, 
SLAC National Accelerator Laboratory, Menlo Park, CA, 94025, USA}
\altaffiltext{5}
{Astronomy Department, New Mexico State University, Las Cruces, NM, 88003, USA {\tt aklypin@nmsu.edu}}
\altaffiltext{6}
{Department of Physics, University of California, Santa Cruz, CA 95064, USA {\tt joel@ucsc.edu}}


\begin{abstract}
  We present a new measurement of the mass of the Milky Way (MW) based
  on observed properties of its largest satellite galaxies, the
  Magellanic Clouds (MCs), and an assumed prior of a \LCDM\ universe.
  The large, high-resolution Bolshoi cosmological simulation of this universe
  provides a means to statistically sample the dynamical properties of
  bright satellite galaxies in a large population of dark matter
  halos.  The observed properties of the MCs, including their circular
  velocity, distance from the center of the MW, and velocity within
  the MW halo, are used to evaluate the likelihood that a given halo
  would have each or all of these properties; the posterior PDF for
  any property of the MW system can thus be constructed.  This method
  provides a constraint on the MW virial mass, $\MassEstimate \Msun$
  (68\% confidence), which is consistent with recent determinations
  that involve very different assumptions. In addition, we calculate
  the posterior PDF for the density profile of the MW and its
  satellite accretion history.  Although typical satellites of
  $10^{12} \Msun$ halos are accreted over a wide range of epochs over
  the last 10 Gyr, we find a $\sim$\accTimeProb\ probability that the
  Magellanic Clouds were accreted within the last
  Gyr, and a \SimultaneousAccFraction probability that they were accreted together.
\end{abstract}

\keywords{Galaxy: formation, fundamental parameters, halo --- galaxies: dwarf, Magellanic Clouds, evolution --- dark matter}


\section{Introduction}

The contents of the Milky Way (MW) Galaxy and the satellites that fall
under its dynamical spell provide a unique testbed for theories of
galaxy formation and cosmology. Detailed observations of resolved
stars, including proper motions, allow the mass distribution of the
galaxy to be measured with ever higher precision
\citep[e.g.][]{Kallivayalil06a, Piatek08}. Satellite galaxies within
the MW have been detected with luminosities three orders of magnitude
smaller than in external galaxies \citep[e.g.][]{Belokurov07}. In
addition, determining the detailed phase space distribution of the MW
galaxy is critical for interpreting the results of experiments to
directly or indirectly detect particle dark matter
\cite[e.g.][]{StrigariTrotta, Vogelsberger09, Lisanti10, Kuhlen10}.  A
full understanding of the MW's place in the Universe requires not only
detailed knowledge of its mass distribution and formation history, but
also a sense of how this one well-studied system fits into the full
cosmological context.

A variety of methods have been used to put limits on the MW mass,
ranging from stellar dynamics and dynamics of satellites
\citep{Klypin02, Battaglia05, Dehnen06, Smith07, Xue08, Watkins10,
  Gnedin10} to the dynamics of the local group \citep{LiWhite08}.  For
example, \cite{Battaglia05} and \cite{Xue08} performed a Jeans
analysis of measurements of the radial velocity dispersion
profile from satellite galaxies, globular clusters, and blue horizontal
branch halo stars to estimate the MW radial density profile.
\cite{Smith07} used measurements of high-velocity halo stars to
estimate the MW escape speed assuming an NFW profile.  \cite{LiWhite08} took a
complementary approach, using the relative position and velocity of
the MW and M31, along with the age of the universe, to infer properties of the obits of the MW-M31 system, which provides constraints on the total mass.  

Similarly, there have been a range of studies on the dynamical state
of the Magellanic Clouds (MCs).  Until recently, the standard picture
was that the MCs were objects that have been orbiting the MW for some
time \citep{Murai80, Gardiner94}.  This picture was motivated in part
by the presence of the Magellanic Stream, a filament of gas extending
$150^\circ$ across the sky.  Because it is apparently spatially and
chemically associated with the Magellanic Clouds, it has often been
interpreted as a tidal tail and taken as an indication that the
satellites have been around for several Gyr \citep[see, e.g.,
][]{Lin82,Connors06}.  This picture has recently come under fire, largely as
a result of detailed measurements of the three-dimensional velocity of
the MCs: they are observed to have high velocities not aligned with
the Magellanic Stream, indicating that they are not in virial
equilibrium \citep[and suggesting alternative formation methods for
the Magellanic Stream; see ][]{Besla07, Besla10}.  Similarly, there is a
growing consensus that the MCs accreted as a group: evidence for this
comes from both their proximity in phase
space~\citep{Kallivayalil06b}, and the result that simulated subhalos
in general tend to accrete in groups \citep{DOnghia08}.

In this work, we take a new approach to measure the mass and assembly of the
MW.  N-body simulations of dark matter structures in a \LCDM\ universe
have been very successful at reproducing the observed clustering of
galaxies \citep[e.g.][]{Kravtsov04,Conroy06,TrujilloGomez10}.  In two companion
papers, we show that the full probability distribution (PDF)
for the number of bright satellites around MW-luminosity hosts
predicted by high resolution numerical simulations \citep{Busha10} is
in excellent agreement with measurements from the SDSS \citep{Liu10, Tollerud11}.
This provides evidence that such cosmological simulations realistically
represent galaxy halos and their satellites, and that they sample the
underlying probability distribution for the properties of halos and
subhalos in our Universe. These simulated halo
catalogs therefore constitute a {\it highly informative prior PDF for the
  parameters of any particular galaxy system}.  In this work we show
that combining this prior with basic data about the two most massive
MW satellites --- their masses, velocities, and positions --- provides
interesting constraints on the MW mass, the distribution of mass
within the MW system, and the system's assembly history.  Although we
find that the simulation used here provides only a relatively sparse
sampling of the underlying PDF, this is the first time that the
statistics to study the MW system in this way have been available at
all.  As high-resolution cosmological simulations probe ever larger
volumes, this approach will have increasing power and applicability.


\section{Simulations}\label{sec:sims}

Statistical inference from halo dynamical histories requires a large,
unbiased set of dark matter halos which samples the full range of
cosmologically appropriate formation scenarios.  Here, we use halos
from the Bolshoi simulation \citep{Klypin10}, which modeled a
250~$\hmpc$ comoving box with $\Omega_m = 0.27$, $\Omega_{\Lambda} =
0.73$, $\sigma_8 = 0.82$, $n=0.95$, and $h = 0.7$.  The simulation
volume contained $2048^3$ particles, each with mass $1.15 \times 10^8
\hinv\msol$, and was run using the ART code \citep{Kravtsov97}.  Halos
and subhalos were identified using the BDM algorithm \citep{Klypin97};
see \cite{Klypin10} for details.  One unique aspect of this simulation
is the high spatial resolution, which is resolved down to a physical
scale of 1~$\hinv \kpc$.  This improves the tracking of halos as they
merge with and are disrupted by larger objects, allowing them to be
followed even as they pass near the core of the halo.  The
  resulting halo catalog is nearly complete for objects down to a
  circular velocity of $\vmax = 50$ \kms.  While the overall
  simulation suffers from incompleteness in satellites at the $\sim$
  20\% level at 50\kms, the incompleteness is strongly dependent on
  host mass.  Host halos with $\mvir = 0.5-3\times 10^{12}\hinv\msol$
  seem to be missing fewer than $\sim 10\%$ of their subhalos.  
  Because we are primarily interested in these lower mass objects, the
  amount of incompleteness is small and can be ignored.

The large volume probed results in a sample of
2.1~million simulated galaxy halos at the present epoch, including
more than 100,000 halos massive enough to host at least one resolved
subhalo.  We can increase this number further by considering halos
identified at different epochs to be independent objects
representative of local systems: we use halos from 60 simulation
snapshots out to redshift 0.25.  Throughout, we define ``hosts'' as
halos that are not within the virial radius of a larger halo, and
``satellites'' as any object within 300~kpc of a host.  This
value is chosen to be roughly half the distance between the MW and
M31.  Note that our final results are largely independent of this
choice, because we will later impose more stringent
requirements on the distance from the
satellites to the center of their host halo.  


\section{Observations}\label{sec:data}

We now consider the massive subhalo population of the MW that is
modeled by the Bolshoi simulation: objects with $\vmax > 50$ \kms.
The two brightest MW satellite galaxies, the LMC and SMC, have both
been measured to have maximum circular velocities $\vmax \gsim 60$
\kms\ with magnitudes $M_V = -18.5$ and -17.1, respectively
\citep{vanderMarel02, Stanimirovic04, vandenBergh00}.  The next
brightest satellite is Sagittarius, some 4 magnitudes dimmer, with
$\vmax \sim 20$ \kms \citep{Strigari07d}.  Similar constraints, albeit
with larger error bars, can be made for the other bright classical
satellites.  The census of nearby objects brighter than $M_V \approx
-8$ should be complete well beyond the MW virial radius \citep{Walsh09,
  Tollerud08}.  It is therefore a robust statement that the MW has
exactly two satellites with $\vmax \gsim 50$\kms.

Applying this selection criterion to the Bolshoi catalog, we find
36,000 simulated halos that have exactly two satellites with $\vmax >
50$\kms. These $\Nsubs = 2$ systems represent a first attempt at
finding simulated halos that are analogs of the MW in terms of massive
satellite content.  What other properties of the LMC and SMC might
provide information on the mass and assembly history of the MW system?
Repeated observations over many years with {\it HST} and ground-based
spectroscopy have given us excellent limits on both the 3D position
and velocity of both objects.  Indeed, both of these properties are
significantly better constrained than are the circular velocities of
these objects.  We select simulated objects to be MC analogs based on
$v_{\rm max}$, the maximum circular velocity of the object, $r_0$, the
distance of the object to the center of the MW, 
and $s$, the total speed of the object relative to the MW.  These are 
summarized in Table~\ref{table:properties}.  In
order to be conservative with the uncertainties to account for
systematics, we multiply the published formal errors in the 
speed by a factor of two when looking for MC
analogs in Bolshoi (included in the errors shown in Table
\ref{table:properties}).  This increase is necessary to bring the velocity 
measurements of the SMC by \cite{Kallivayalil06} and \cite{Piatek08}
into agreement.  

Note that there is a slight discrepancy between the
observational measurements of satellite $\vmax$ and the
simulations: while observations measure the circular velocity curve
for all components of the halo --- dark matter and baryons --- the
Bolshoi simulation models only the dark matter content.
\cite{TrujilloGomez10} showed that ignoring the baryonic component
may cause a $\sim 10\%$ over-estimate in the measurement of $\vmax$ for
MC-sized objects.
Since this correction is smaller than the error bars on the
observations, we choose to ignore it.  Additionally, there is some
disagreement in the literature as to the allowable upper limit for
$\vmax$.  In particular, some estimates for the LMC are in excess of
100 km/s \citep{Piatek08}, well above our adopted 80km/s $1\sigma$
upper limit.  However, due to the rapidly declining abundance of
such massive subhalos \citep[see, e.g.,][]{Busha10}, our analysis is
highly insensitive to changes in the upper error bar on $\vmax$.  Finally,
because of resolution effects, there is a radial dependence to the
incompleteness of satellite halos, with the large density contrast and limited
force resolution making it difficult to identify subhalos closer to the center
of their hosts.  However, this incompleteness does not appear to correlate
with host mass, so we do not expect it to bias our results.

\begin{figure*}
\begin{center}
\resizebox{0.47\textwidth}{!}{\includegraphics{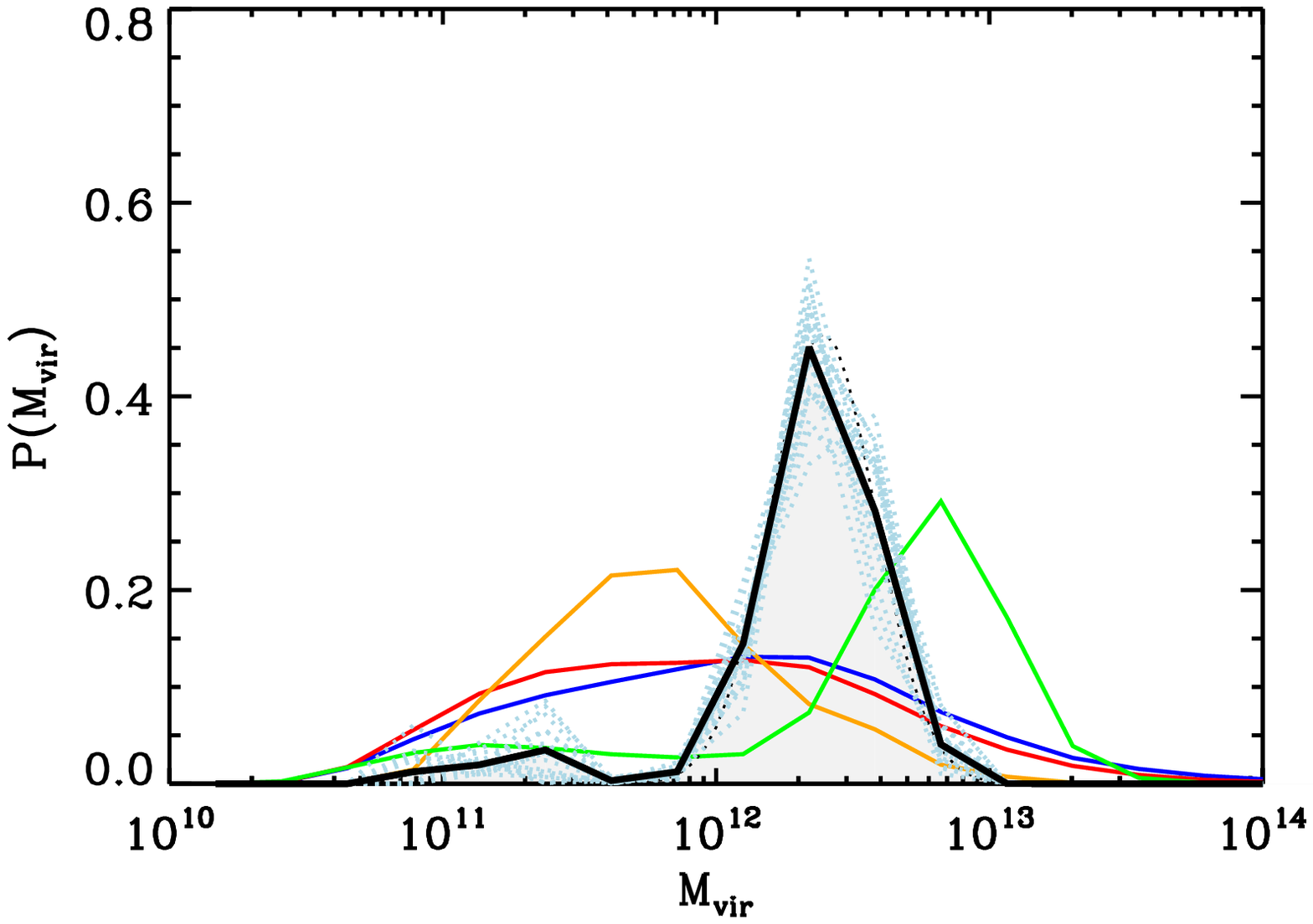}}
\resizebox{0.47\textwidth}{!}{\includegraphics{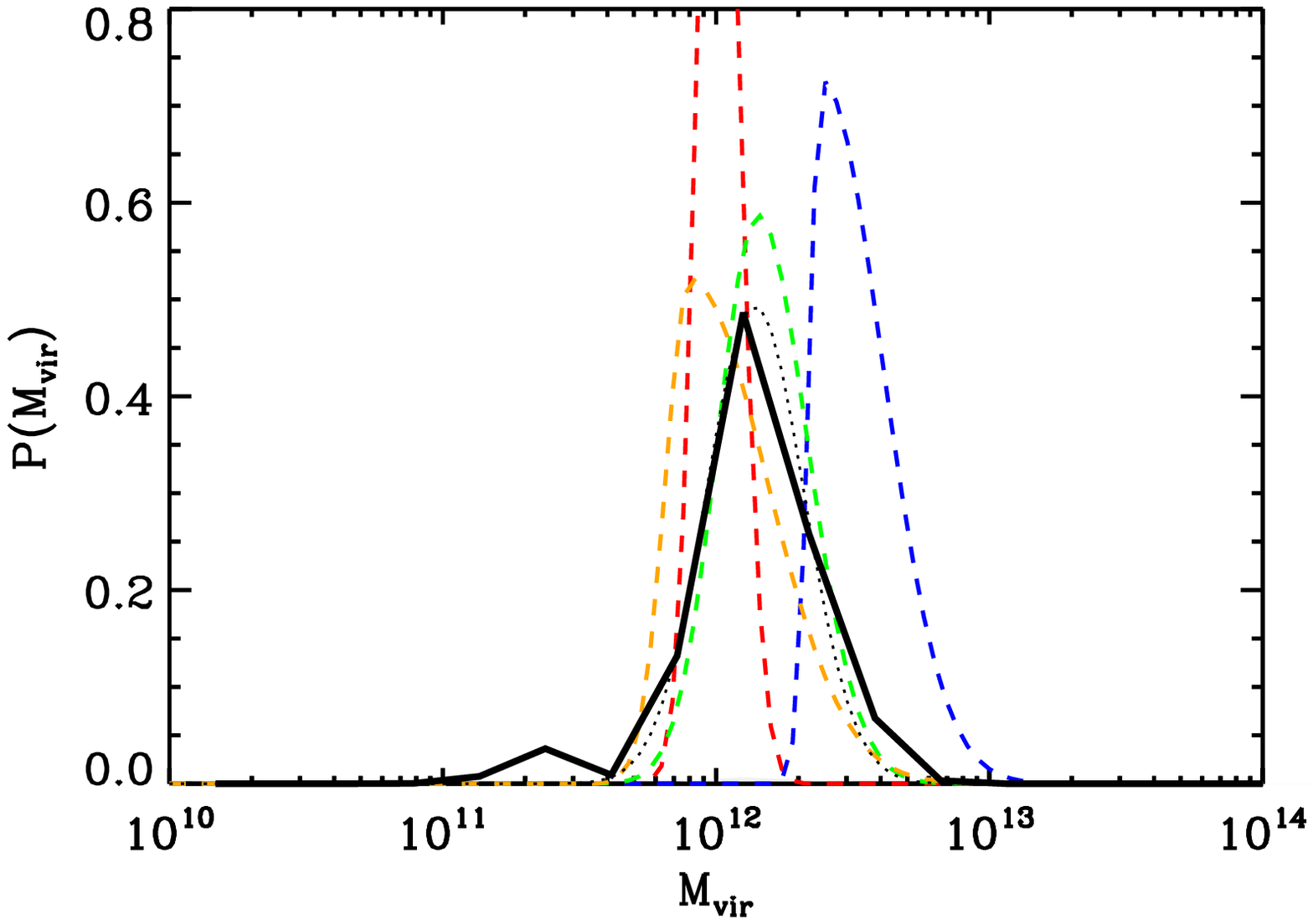}}
\hfill
\end{center}
\caption{ {\em Left:} The MW mass inferred from the properties of its
  two most luminous satellites, the Magellanic Clouds.  Lines show
  posterior PDFs (weighted histograms of Bolshoi halos) 
  given information about (a) the existence of exactly two
  satellites with $v_{\rm max} > 50$ km s$^{-1}$ (blue), 
  (b) the
  maximum circular velocities $v_{\rm max}$ of the two satellites
  (red), (c) the distance of each satellite from the center of the MW
  (orange), (d)
  the speed of each satellite (green), and (e) all of these properties
  simultaneously (black).  The combined properties 
  define a sample of ``satellite analogs'' and
  give $\MMW =
  \MassEstimate \Msun$ (68\% confidence). The dotted line shows a
  lognormal fit to this distribution, with parameters $\log_{10}{\MMW}
  = \logMassEstimate$, and the dashed gray lines show the effect of 
  bootstrap resampling: the uncertainty on the mean of the
  distribution from 
  ``sampling noise' in the catalogs was found in this way to be just 
  $0.03$~dex. \\
  {\em Right:} Comparison with various estimates for the mass of the
  MW from the literature.  Dashed lines show results from: (a) the
  radial velocity dispersion profile \citep{Battaglia05} (orange); (b)
  the escape velocity from halo stars \citep{Smith07} (red); (c) SDSS
  blue horizontal branch stars \citep{Xue08} (green), and (d) the
  timing argument \citep{LiWhite08} (blue).  We assume lognormal error
  distributions, with asymmetric errors given by the quoted upper and
  lower confidence limits.  The solid (black) line shows the posterior
  PDF for the MW mass from our satellite analogs, 
  and the dotted line again shows thee lognormal fit to this distribution.}
\label{fig:mass}
\end{figure*}

\def\arraystretch{1.4}
\begin{deluxetable}{cccc}[ht!]
\tabletypesize{\footnotesize}
\tablecaption{\label{table:properties}
Observed properties of the LMC and SMC.}
\startdata
\hline
               & LMC          & SMC          & Reference          \\
\hline
\hline
$\vmax$ [km/s] & $65 \pm 15$  & $60 \pm 15$  & vdM02, S04, HZ06   \\
$r_0$ [kpc]    & $50 \pm 2$   & $60 \pm 2$   & vdM02              \\
$s$ [km/s]$^*$     & $378 \pm 36$ & $301 \pm 104$ & K06
\enddata
\tablecomments{
For a given satellite, $\vmax$ is its 
estimated maximium circular velocity,
$r_0$ its estimated distance from the Galactic center, and $s$ its
estimated speed relative to the Galactic center. References are:
vdM02~=~\citet{vanderMarel02};
S04~=~\citet{Stanimirovic04};
K06~=~\citeauthor{Kallivayalil06a}~(\citeyear{Kallivayalil06b},\citeyear{Kallivayalil06a})
HZ06~=~\citet{Harris06}. \newline
$^{*}$~Errors on $s$ have been increased
relative to the published values for conservatism (see text).}
\medskip
\end{deluxetable}


\section{Inference}\label{sec:prob}

The Bolshoi halo catalog is large enough for some of its members to resemble
the Milky Way system, with regards to its two most massive satellites. By
weighting the Bolshoi halos according to how closely their satellites'
properties match those of the MW's, we can infer the mass and assembly history of the
MW system. In this section we explain how this works, and derive the required
weights from probability theory.

\subsection{Observational constraints and the likelihood function}

As outlined above, the Bolshoi halo catalog can be thought of as a set of
samples drawn from an underlying probability distribution. 
Each halo is
characterized by a set of $m$~parameters, $\pars$, which includes the total
mass of the halo and the properties of its subhalos, such as their masses,
positions, and velocities: 
$x = \left( M_{\rm vir}, \{v_{\rm max},r_0,s,\ldots\}_{\rm LMC,SMC}\right) $.  
Given no observational data, the set of Bolshoi
halos provides a reasonable characterization of our prior PDF for the
parameters of the Milky Way system,~$\pr(\pars)$. For example, if we were
asked to guess the mass of the MW halo, we would do  much worse by drawing a
random number from some wide range, than we would do by drawing one Bolshoi
halo at random from the catalog.

However, we do have some observational data: we would therefore  like to know
the posterior PDF for the parameters of the Milky Way system,  given this
data~$\data$. Bayes' theorem shows this to be:
\begin{equation}
  \pr(\pars|\data) \propto \pr(\data|\pars) \pr(\pars)
  \label{eq:bayes}
\end{equation}
The first term on the right hand side is the joint likelihood function  for
the parameters, written as a function of the data. Given a particular
parameter vector $\pars$, we can compute the value of this likelihood in the
usual way, given assumptions about the error distributions of the
observational data.  Our data $\data$ and their uncertainties $\sigma$  are
given in Table~\ref{table:properties}: $\data = \left( v_{\rm max}^{\rm
obs},r_0^{\rm obs},s^{\rm obs}\right)$ (for each MC). We use the
superscript ``obs'' to make clear the distinction between the (constant) measured
values, and the corresponding (variable) parameters of the model, which are
the properties of the Bolshoi halos. We interpret the errors on the
observational data  as being Gaussian-distributed, such that for each datum
$\datai$ its likelihood function $\pr(\datai|\pars)$ is a Gaussian, with mean
and standard deviation listed in Table~\ref{table:properties}. 

Since the $N=6$ observations were made independently 
by different groups using different techniques, 
the joint likelihood is just a simple product:
\begin{eqnarray}
  \pr(\data|\pars) &=& \prod_i^N \pr(\datai|\pars), \\
                   &=& \prod_{\rm LMC,SMC}
                       N(v_{\rm max}^{\rm obs}|v_{\rm max},\sigma_{v}^2)\\
                   & &   \;\;\;\;\;\;\;\;\;\;\;\;\;\;\;\;\;\;\;\;\;
                         \times N(r_0^{\rm obs}|r_0,\sigma_{r}^2)
                         \times N(s^{\rm obs}|s,\sigma_{s}^2), 
\end{eqnarray}  
where 
\begin{equation}
  N(y|\mu,\sigma) = \frac{1}{\sigma\sqrt{2\pi}}
                   \exp\left[ -\frac{\left(y-\mu\right)^2}
                                    {2\sigma^2}           \right]
\end{equation}  
is the Gaussian probability density in $y$ with mean $\mu$ and variance 
$\sigma^2$.

Conditional on the values of the parameters, the data measurements are
independent, and thus the likelihood function factors into the product of N=6
terms.
In the prior, the dynamical parameters and halo properties themselves
are correlated, and not independent, because of the simulation physics.
We have a measurement of satellite distance,
$r_0^{\rm obs}$, and an independent measurement of satellite velocity 
$s^{\rm obs}$: the probability calculus makes it very clear that the PDFs
for these two observed quantities 
can be multiplied together, even
though the corresponding underlying \emph{parameters} $r_0$ and $s$ are not
independently distributed. 
(We discuss correlations between the measurements of $r_0$ and $s$ below.)\,\,
The dynamics of satellites around halos is such
that their positions and velocities are correlated according to the orbits
they are on. This is a good thing: it means that by measuring one we can learn
about the other! In fact, it is this very interdependence that will allow us
to infer the MW halo mass given measurements of its satellite properties. All
the correlations between the model parameters are correctly taken into
account by drawing samples from the joint prior PDF -- that is, by using the
Bolshoi catalog.

\subsection{Importance Sampling}

With the likelihood function in hand, we can now calculate the posterior PDF
for the MW system parameters given our data.  
We have the prior PDF $\pr(\pars)$  in the form of a set of~$n$
samples drawn from it; this is actually a very convenient characterization, 
and will allow us to derive an equally convenient characterization of the
posterior. 
We compute posterior estimates using importance sampling (see
\citeauthor{MacKay}~\citeyear{MacKay} for an introduction, and papers
by \citeauthor{L+B02}~\citeyear{L+B02}  and
\citeauthor{Suy10}~\citeyear{Suy10} for example applications in astronomy). 
In general, this technique involves generating samples from an importance 
sampling function, which  are then weighted when computing integrals over the
target PDF.  In this paper, we choose the importance sampling function to be
the prior PDF, so that the importance weights are proportional to the
likelihood; this allows us to compute integrals over the posterior PDF, as
shown below.

These integrals include mean parameter values, confidence intervals and so on;
a histogram of sample parameter values  is a representation of the
marginalized distribution for that parameter, and is also a set of integrals
(counts of samples in bins). 

In general, integrals over the
posterior PDF can be written as follows:
\begin{eqnarray}
  \int f(\pars)\, \pr(\pars|\data)\, d^m\pars 
      &=&  \frac{\int f(\pars)\, \pr(\data|\pars)\, \pr(\pars)\, d^m\pars}
                {\int            \pr(\data|\pars)\, \pr(\pars)\, d^m\pars} \\
      &\simeq& \frac{\sum_j^n f(\pars_j)\, \pr(\data|\pars_j)}
                    {\sum_j^n       \pr(\data|\pars_j)}
\end{eqnarray}  
where in the first line we have substituted for the posterior PDF using
equation~\ref{eq:bayes}, and in the second line we have approximated 
the integrals
with sums over samples drawn from the prior PDF. The denominator in each case
is a normalization constant. $f(\pars)$ is any function 
of interest to be integrated: for example, $f(\pars) = \pars$ gives the
posterior mean parameters.  
The highest importance samples
correspond to halos that most resemble, in terms
of its bright satellite properties, the halo of the MW.

\subsection{Assumptions}

In this subsection, we examine the assumptions we have made in more
detail.

\subsubsection{Data covariance}

As noted above, the constraints on $r_0$ and $s$ are nearly, but not quite,
  independent.  Working in galactic coordinates and following the method
  outlined in \cite{vanderMarel02}, the separation between the MCs and the
  center of the MW, $r_0$, is given by the relation
\be
r_0^2 = \rsol^2 + {\bf r}_e^2 + 2(\rsol \cdot {\bf r}_e),
\label{eq:r0}
\ee
where $\rsol$ is the location of the center of the galaxy and $\bf{r}_e$ location of the MCs, both relative to the sun.  Similarly, the speed, $s = |{\bf s}|$, for the MCs is given by
\be
{\bf s} = \vsol + (\mu_N {\bf u}_N + \mu_W {\bf u}_W)r_e + v_{los} {\bf u}_{los},
\label{eq:s}
\ee
where $\vsol$ is the relative motion between the sun and the center of the
galaxy, $\mu_N$ and $\mu_W$ are the measured north and west transverse
components of the proper motion of the MCs on the sky, $\bf{v}_{los}$ the
measured line of sight velocity of the MCs, and $\bf{u}_N, \bf{u}_W$,
$\bf{u}_{los}$ are the unit vectors defining the (north and west) transverse
and radial directions of motion of the MCs relative to the sun in galactic
coordinates.  

As can be seen in equations \ref{eq:r0} and \ref{eq:s}, $r_e$, the distance
from the Sun to the MCs, is a necessary measurement for determining both $r_0$
and $s$, hence these measurements are not fully independent.  In order to
determine the degree of dependance, we calculate $\rm{COV}(r_0,s)$, the
covariance between these variables, and and show that this is significantly
smaller than the measurement errors on the properties, 
\be
{\rm COV}(r_0,s) \ll \sigma_{r_0}\sigma_s.
\ee
The covariance can be explicitly written as
\begin{eqnarray}
{\rm COV}(r_0,s) &=& {dr_0 \over dr_e}{ds \over dr_e}\sigma_{r_e} = \nonumber\\
{1\over \sqrt{r_0}}(r_e + {\rsol \cdot {\bf r}_e \over r_e}) &\times& {1 \over 2 \sqrt{s}}(\mu_W^2 + \mu_N^2)
\times \sigma_{r_e}
\end{eqnarray}
Using values reported in \cite{vanderMarel02} and \cite{Kallivayalil06b}, we
measure COV($r_0,s$) = 6.52 and 6.86 for the LMC and SMC, respectively.  In
both cases, this at least an order of magnitude smaller than the product
$\sigma_{r_0}\sigma_s = $ 72, and 208, allowing us to treat the measurements
of $r_0$ and $s$ as independent.

\subsubsection{Simulation time resolution}

The observational constraints on the positions of the MCs are actually
tighter by a factor of two  than the resolution of our simulations. The
available Bolshoi outputs are such that a satellite with the LMC's  radial
velocity of $\simeq90$~km/s \citep{Kallivayalil06a} will travel roughly 4~kpc
between sequential snapshots, making this the effective uncertainty on the
radial position of the simulated halos. We increase the size of the positional
uncertainties accordingly, when calculating the likelihood function for the MC
positions.

\subsubsection{Importance sampling failure modes}

Typically there are two ways in which importance sampling can fail. The
first is that the sampling function does not cover the domain of the target
PDF, leaving parts of the PDF unsampled. In our case, we assert that
the Bolshoi simulation does sample the
parameter space, in that it contains (in large numbers) halos and subhalos
with masses, positions and velocities sufficiently close to the members of the
MW system to make inferences meaningful. The second failure mode is that too
few samples are drawn in the high importance volume, leading to estimates
dominated by a small number of high importance samples. As we shall see, this
``sampling noise'' does lead to additional uncertainty on our inferences.
Because of the very tight observational constraints on the
properties of the MCs, relatively few Bolshoi halos receive significant
importance -- we can only partially compensate for this by searching
through multiple simulation outputs to improve the statistics of our sample.
In Sections~\ref{sec:results:mass} and~\ref{sec:results:history} below, 
we estimate the uncertainty, due to sampling noise, on each of our
estimates by bootstrap resampling.

\subsubsection{The effect of the Galactic disk}

Finally, the lack of treatment of baryons in Bolshoi introduces a systematic
error.  The more concentrated mass in the stellar disk at the halo center
should increase the speed of the satellites orbiting their hosts at small
radii.  We can estimate the impact of this by artificially increasing the
total velocity of our simulated satellites by the circular velocity due to the
stellar disk of the Milky Way, $v_{circ} = \sqrt(GM_*/r_{sat})$, where $M_* =
6 \times 10^{10}\msol$\citep{Klypin02}.  This correction is admittedly
simplistic, but is also small. To conservatively quantify the residual
systematic error associated with this correction, we take the difference in
mass estimates with and without the above correction. This gives us an
approximate upper limit on the size of the two-sided systematic error due to
baryon physics.


\section{The Mass Distribution of the Milky Way}
\label{sec:results:mass}

Figure~\ref{fig:mass} (left panel) shows the posterior PDF for the MW
halo mass, computed by weighting every Bolshoi halo by the probability
that its bright satellite population looks like that of the MW in
various ways.  Some observations provide significantly more
information about the host halo mass than others: the distance to and
speed of the MCs are most constraining, while the maximum
circular velocity of the LMC and SMC provide almost no information
(largely due to their measurement errors).  Note that it is the {\it
  combination} of datasets that is most important, as internal
degeneracies are broken, i.e., there is a high degree of covariance
between positional and kinematic properties.  The combination of all
datasets gives $\MMW = \MassEstimate \Msun$ (68\% confidence) and a
virial radius $r_{vir} = 250^{+60}_{-30}\kpc$.  
The systematic errors are estimated by repeating this process with and
without the baryon correction discussed at the end of Section 4.

How many halos contribute to this inference? We find 114 $1\sigma$
matches (systems that are within an average of $1\sigma$ from observations of
the MCs in the 6 properties listed in Table~\ref{table:properties}) and nearly
400 $2\sigma$ matches in the 60 snapshots. However, many more contribute
statistically.  The effective number of halos contributing to the posterior
can be estimated as $N_{\rm eff} = N / (1 + {\rm Var}(w))$~\citep{Neal01},
where $w$ is the normalized importance of each halo and $N$, in our case, is
$1.71\times10^6$ (total number of halos over all 60 snapshots).  We
find $N_{\rm eff} = 10,051$.

While this is a healthy number of samples to compute statistics with,
the low {\it fraction} of
halos that contribute is driven largely by the combination of the total speed, $s$, and radial
position, $r_0$, constraints.  This gives additional support to the idea that
the MCs are currently at-or-near pericentric passage on their orbits around
the Milky Way, since any object on an elliptical orbit spends such a small
amount of time near pericenter. 
{Figure~\ref{fig:mass} shows the
effects of the sampling noise on the inference of the MW mass: the faint blue-grey
curves show 25 bootstrap-resampled PDFs. The additional uncertainty on the
central value is 0.03 in $\log_{10}(\MMW)$ -- this is included this in the quoted error
bars above, which were estimated from the sum of the resampled PDFs.}

Figure~\ref{fig:mass} (right panel) compares our result with previous
MW halo mass estimates from the literature.  The LMC and SMC
properties lead to a halo mass that is in excellent agreement with the
dynamical estimates in the literature.  In particular, our results are
in near perfect agreement with the most recent stellar velocity
measurements \citep{Xue08}, with similar error bars.

\begin{figure}[!ht]
\begin{center}
  \resizebox{0.47\textwidth}{!}{\includegraphics{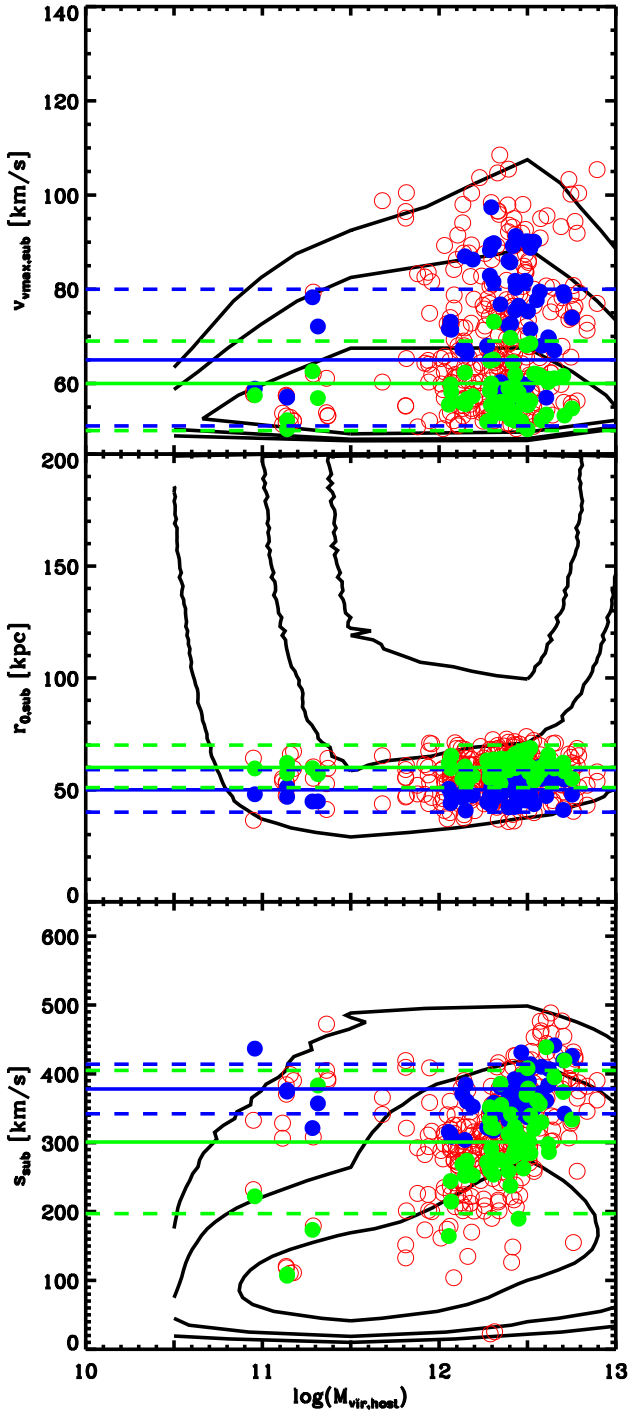}}
  \caption{Relationships between $M_{vir,host}$ and various satellite
    parameters:  $\vmax$ (top left), $r_0$ (top right), and $s$ (bottom left).  In all panels, the black contours denote
    the regions containing 68, 90, 95, 98, and 99\% of our prior distribution,
    all satellites in a host with $N_{sub} = 2$.  
    The open red and filled circles denote the satellites
    for 2- and 1-$\sigma$ halos, those that contribute most to the posterior
    PDF.  Blue and green lines show the observed
    values and $\pm 1-\sigma$ uncertainty for each property,
    for the LMC and SMC respectively.  Similarly, the blue and green circles denote the values for our identified LMC and SMC analogs.  }
\label{fig:satellite_properties}
\end{center}
\end{figure}

Throughout the paper, we refer to the collection of hosts weighted by
the likelihood of the
$\vmax, r_0$, and $s$ of their satellites as ``satellite analogs'' of the
MW.  Note that this does not imply that we have selected a specific subset of
hosts.  Rather, we have taken {\em all} hosts with $N_{subs} = 2$ and weighted
each object by its satellite properties.  It is this sample of weighted objects
that defines our satellite analogs.  It is worth noting, however, while we
formally use all 35,000 $N_{subs} = 2$ halos in Bolshoi, 
the $\sim 500$ 1- and 2-$\sigma$ halos provide more than 95\% of the
total weight.

Because we want to further understand what impact the MCs have on other other
halo properties,  we also define ``mass analogs'' to be the set of halos randomly
drawn from the mass PDF of the satellite analogs. Thus, the mass analogs have
the same PDF of virial masses as the black line of Figure
\ref{fig:mass}, but no constraints  on their satellite
properties.  Comparison between our satellite analogs and mass analogs
allows us to disentangle impacts on the system due to the satellite
properties from those due to the particular mass range probed by our
satellite analogs.  

When interpreting Figure~\ref{fig:mass}, it is also helpful to consider the
full dependence of the predicted satellite properties on the mass of
the host halos. This is shown in
Figure~\ref{fig:satellite_properties}, which compares the
satellite parameters  $v_{max,}, r_{0}$, and $s$ 
with the $\mvir$ of their hosts.  The plots show 
contours containing 68, 90, 95, 98, and 99\% of our 
prior probability (that is, all hosts with
exactly two satellites, black lines), 
as well as the exact locations of our
2-$\sigma$ (open red circles) and 1-$\sigma$ hosts
(filled circles).  Also plotted are the \emph{observed} 
properties of the
LMC and SMC (blue and green lines).  These plots highlight a number of
important trends.  First, we can see that the MCs are atypical
subhalos in most regards.  The LMC in particular is roughly a 2-$\sigma$ outlier in each of these properties.  
Second, correlations between parameters can
be seen, such as the that between speed and host mass, which
shows explicitly how the high \emph{observed} 
speed of the satellites skews the
posterior PDF towards higher mass hosts.  
Additionally, it is interesting
to note that, while rare, there are a few objects with $\mvir <
10^{11}\msol$ with satellites that are well matched to the MCs.  These
satellites are fast moving, massive subhalos roughly 50 and 60 kpc 
from their host centers -- well within their host virial radii, but 
not energetically bound to their hosts, and hence merely transient events.
Finally, while it can be seen that
observations of $s$ provide the most stringent constraints
individually, these plots further emphasize that it is the combination
of observed properties that is necessary to place tight constraints on
$\MMW$.

We can also determine the impact of the MCs on the internal mass distribution of a halo by comparing the density profiles for our satellite and mass analogs.  This gives us a handle on how typical the MW is for a halo of its mass.  
We find that the presence of the MCs has only a modest impact on the
halo concentration, but can impact density profiles significantly 
in other ways.  In particular, when looking at
the mass enclosed within a fixed 8 kpc (the distance from the sun to the center of the galaxy),
satellite analogs have a 60\% higher central density than mass
analogs.  This tells us that the MCs are correlated with a more
strongly peaked inner dark matter distribution.  For the full density profile
(which includes contributions from substructures), however, satellite analogs have concentration $c_{vir} = \cEstimate$,
a little less than 1-$\sigma$ higher than $c_{vir} = \cpopEstimate$ for the mass analogs (here, $c_{vir} = r_{vir}/r_s$).  While this still implies a correlation between the presence of the MCs and a more peaked mass distribution, it shows that the impact is much weaker at larger radii than it is for the central core.  


\section{The Assembly of the Milky Way Halo}
\label{sec:results:history}

\begin{figure}[t!]
\resizebox{0.47\textwidth}{!}{\includegraphics{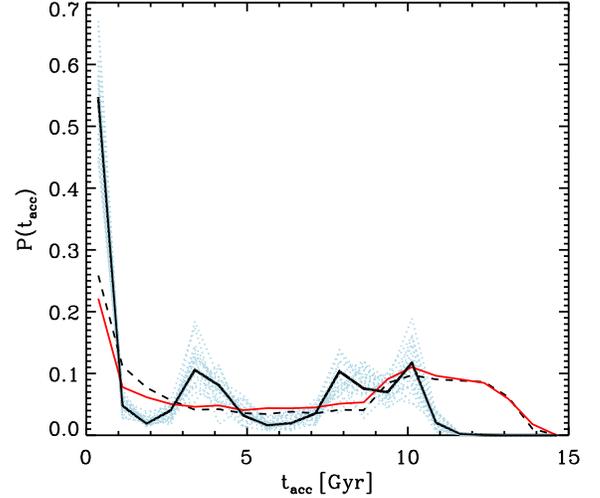}}
\caption{Posterior distribution of satellite accretion times, from the MW
  satellite analogs (black), from hosts with exactly two subhalos
  (dashed), and from MW mass analogs (red).  Selecting hosts with
  MC-like satellites strongly weights the distribution towards recent
  accretion.
   The blue-gray dotted lines show the distributions for 25 random MW satellite analogs drawn from our bootstrap analysis.}
\label{fig:accretion_times}
\end{figure}

We now turn our attention to the assembly history of the Milky Way.
We use the same importance-sampled halos from the previous section to
infer the MW assembly history, in the same way as we inferred its mass
and density profile.  Figure~\ref{fig:accretion_times} shows the
distribution of accretion times for these satellite analogs (where $t_{acc}$ is the time since the subhalo first came within 300 kpc of the host); we also
show the accretion time PDF for all hosts with $\Nsubs = 2$, and for
the mass analog systems defined above.  The mass analogs (red line)
clearly show two populations.  The first consists of
halos whose subhalos were accreted at high redshift, when the
host halo was in its exponential growth phase, which suppresses
tidal stripping of the subhalos \citep{Wechsler02, Busha07}.  The second
population consists of halos with recently accreted objects
that have not had enough time to undergo significant tidal disruption.
The relative size of these populations changes when we apply the
observational likelihoods.  
Requiring that a host has $\Nsubs = 2$ massive
subhalos (with unconstrained speeds and distances, dotted line) 
has little impact.  However, for satellite
analogs (black line), the size of the recently accreted
population increases dramatically.  This is primarily driven by the
combined requirement that the satellites have both a high radial
velocity and are close to the center of the halo, and argues that
there is roughly a \accTimeProb\ chance that MCs are recent arrivals, accreted within
the past \accTimeLimit.  
While it is clear that low number statistics are strongly impacting the
 shape of the probability distribution for the accretion time of the  mass
 analogs in Figure \ref{fig:accretion_times}, our bootstrap analysis shows that
 the measurement of \accTimeProb\ of all subhalos accreting within the last
 \accTimeLimit is a robust result that is not sensitive to the small number
 statistics. This is again demonstrated by the gray-blue lines in Figure \ref{fig:accretion_times}, which show the distriubtion of accretion times in 25 or our resamplings.  In all of these cases, there is a strong peak at recent accretion times.  
Our bootstrap analysis shows that
the measurement of \accTimeProb\ of all subhalos accreting within the last
\accTimeLimit is a robust result, that is not sensitive to sampling noise.
Indeed, in 95\% of our resamples, there is a $>65\%$ probability
that the Magellanic clouds were accreted within the past \accTimeLimit.

\begin{figure}[t!]
\resizebox{0.47\textwidth}{!}{\includegraphics{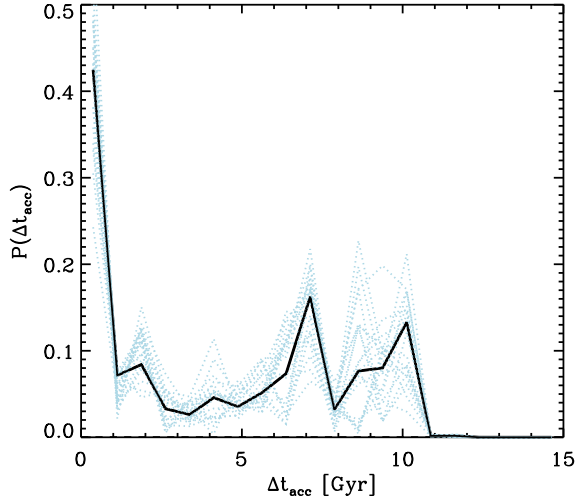}}
\caption{Difference between accretion times for the two most massive
   satellites in MW-like systems.  The solid line represents satellite
   analogs; the dashed line shows all systems with exactly two subhalos.
   The blue-gray dotted lines show the distributions for 25 random MW satellite analogs drawn from our bootstrap analysis.
 }
\label{fig:selected_accretion_times}
\end{figure}

Did the MCs arrive together?
Figure~\ref{fig:selected_accretion_times} shows the difference in
accretion times, $\Delta t_{\rm acc}$, for the two most massive
satellites of all hosts with $\Nsubs = 2$ and for the two satellites
of the satellite analog hosts.  For satellite analogs, the
distribution is strongly peaked towards small $\Delta t_{\rm acc}$,
with roughly \SimultaneousAccFraction\ of satellites having been
accreted simultaneously (within a Gyr of each other).  These
simultaneous accretions correspond very tightly with the recently
accreted population.  
The noise in this plot, and in particular the peak around
$\Delta t_{\rm acc} \approx 8-9$Gyr is driven by a few very well
matched (high weight) halos that had one satellite accrete
within the last Gyr and the other early on during the exponential
buildup phase (see the weaker secondary peak in Figure
\ref{fig:accretion_times}).  This highlights the current level of
noise in our analysis due to our modest statistical size.  We
anticipate that, with better statistics, this high-$\Delta t_{\rm
  acc}$ bump will smooth out, making a smoother distribution that is
even more sharply peaked at $\Delta t_{\rm acc} = 0$.  Still, using out 
bootstrap analysis, the large fraction of objects accreting within a Gyr
 of each other appears to be quite robust, as can be seen by the locus of blue-gray lines in Figure \ref{fig:selected_accretion_times}.  For
comparison, halos with $\Nsubs = 2$, shown as the dashed line, have a
much weaker preference for simultaneous accretion.

Both this
result and that of Figure~\ref{fig:accretion_times} favor a method for
the creation of the Magellanic Stream other than tidal disruption by
the MW.  For example,
the model of \cite{Besla10}, who found good agreement with the
dynamics of the Stream for a model in which it was created by tidal
disruption of the SMC by the LMC before they were accreted as a bound
pair.  
Additionally, the satellites in the satellite-analogs have a high degree
of spatial correlation, with a mean separation of $48 \pm 8 \kpc$ at the
epoch where the simulations best match observations of the MCs, about
$3\sigma$ larger than the observed MC separation of 25 kpc
\citep{Kallivayalil06b}, yet roughly $3\sigma$ closer than expected for two
randomly drawn points at the appropriate radii for the MCs.  This provides 
some support for the notion that the MCs may be a bound pair.
However we do not detect a lower relative speed of the two subhalos in
the satellite analog sample than in the mass analog sample.
The Bolshoi simulation lacks the volume to address this
question more directly: ideally, we would like to further weight our sample by the
observed separation between the two satellites, but this would
reduce the effective number of halos below the sampling noise limit.  
While we cannot say very much about the boundedness of the LMC and SMC pair,
we can make the following point:
that even without using the observed angular separation of the LMC and SMC
as a constraint, we find a high probability of them arriving at the same
time.

\section{Conclusions}  

The advent of high-resolution cosmological simulations which sample
the dynamical histories of large numbers of dark matter halos in a
wide range of environments provides us with a new approach for
determining the properties of individual halos given their
observational characteristics.  Here, we use the observed properties
of the Magellanic Clouds to constrain the mass distribution and
assembly history of the Milky Way.  In comparison to previous efforts
which use detailed observations but a necessarily simplified dynamical
model, our approach uses simple observations and statistical inference
from sampling of a detailed and cosmologically consistent dynamical
model.

Our principal conclusions are:
\begin{enumerate}
\item We infer the MW halo mass to be $\MMW = \MassEstimate \Msun$
  (68\% confidence), in very good agreement with the recent stellar
  velocity measurements \citep{Xue08}, with similarly sized error bars.
\item The MW halo has a slightly higher concentration than is typical
  for its mass: $\cEstimate$, compared to $\cpopEstimate$.  Additionally,
  the density within 8 kpc is 60\% higher for satellite analogs than for
  mass analogs. 
\item Typical bright satellites of halos with MW mass were accreted at
  a range of epochs, generally at high redshift (c. 10~Gyr ago) or
  much more recently (within the last 2~Gyr).  Because of their high speed near the center of the halo, we find a
  \accTimeProb\ probability that the Magellanic Clouds were accreted
  within the last Gyr.  We also find a
  \SimultaneousAccFraction\ probability that the MCs were accreted
  within 1~Gyr of each other.
  \end{enumerate}

This approach clearly allows one to explore a wide range of
additional properties of MW-like halos; however, it places challenging
requirements on the simulations used.  In particular, for MW studies
we are still limited by the relatively small simulation volume
used.  With the constraints used here, we found just one good
fit halo per $\sim$ \Ngoodvolume, emphasizing the large volume required to 
perform this analysis.
As more criteria are applied we will need to sample a larger range of
formation histories and environments drawn from a larger cosmological
volume.  In addition, 
the properties of its smaller satellites are not accessible with our
present resolution.  Pushing forward on both simulated resolution and
volume will be essential to realize the full potential of this
approach.  Additionally, the simulations used here ignore the impact
of baryons on the dark matter distribution.  We have included a simple
model of the stellar disk which is applied to the satellite velocities
to get a handle on the impact of this systematic, but the model is
simplistic and the effects of the baryonic component need to be
further studied.

As we were completing this work, \citet{Boylan-Kolchin10} presented
results from a similar study.  Their principal results regarding the
mass of the MW and the accretion history of the LMC and SMC are in
reasonable agreement with our own, although they favor a somewhat
larger MW mass, $\Mvir \sim (2-3)\times 10^{12}\msol$.  These
  studies differ in the 16 times larger volume simulation used here, the
  cosmology of the simulations, as well as our use of statistical
  inference.  \citet{Boylan-Kolchin10} used a simulation with
  cosmological parameters taken from the 1-year WMAP results, with
  $\Omega_M = 0.25$, $\sigma_8 = 0.9$, and $h = 0.73$, while the Bolshoi
  simulation uses the more recent WMAP7 results and features a lower
  $\sigma_8 = 0.82$, with $\Omega_m = 0.27$ and $h = 0.7$.  Adopting
  the \citet{Tinker08} mass function, we can understand the impact of
  these cosmological differences by considering the abundances of MW
  and MC mass objects.  Differences come from a number of competing effects.
  First, due to cosmological differences, their lower $\sigma_8$ will tend to suppress our mass function, while
  the lower $h$ will increase the halo masses.  In this case, the
  differences in $h$ has the more significant impact, and causes the
  abundances of halos with $\mvir = 1.2 \times 10^{12}\msol$ to be
  suppressed by about 10\% in the Millennium cosmology.  However, the
  more relevant number is the ratio of the number density for our
  selected satellites to their selected hosts.  Adopting the fiducial
  mass for the LMC, 100 times lower than that of the MW, we see
  that the ratio in number densities is $65.0$ in Bolshoi vs. $63.2$
  in Millennium, a suppression of roughly $\sim3\%$.  Because the mass
  function is relatively flat here, reproducing the Bolshoi abundance
  ratio requires that the host mass increases to $3.6\times
  10^{12}\msol$, a correction that brings our results into better
  agreement with theirs.   However, it is also important to note that
significantly different selection criteria for identifying ``MW-like''
objects were employed.  \citet{Boylan-Kolchin10} selected hosts whose
two largest subhalos were within 0.75 $\rvir$ and had similar total velocities and stellar
masses to the MCs using abundance matching to estimate the stellar
content of their simulated halos.  In this work, we select objects
with exactly two subhalos more massive than $\vmax = 50$\kms within a
fixed 300 kpc aperture and then weight our sample according to how
well the subhalos look like the MCs in terms of $\vmax$, position,
radial velocity, and total speed.  These differences likely account
for the tension in the resulting $\MMW$ PDFs.  In particular, their focus on the total speed, $s$, of the magellanic clouds has likely biased their mass estimates for the MW high.  This is shown by comparing the green and black lines in Figure \ref{fig:mass}, which show the likelihood distribution of masses for hosts having two satellites with similar speed to the MCs (green), as well as for hosts with two satellites with similar speeds, masses, and positions (black).  The green distribution presents a much better agreement with Figure 13 of \cite{Boylan-Kolchin10}.  

\acknowledgments MTB and RHW were supported by the National Science
Foundation under grant NSF AST-0908883. MTB recieved additional funding from the Swiss National science Foundation under contract 200 124835/1.  PJM acknowledges the Kavli
Foundation and the Royal Society for support in the form of research
fellowships.  AK and JRP were supported by the National Science Foundation under grant NSF AST-1010033.  We thank Louie Strigari, Brian Gerke, Nitya
Kallivayalil, and Gurtina Besla for useful discussions.  The Bolshoi
simulation was run using NASA Advanced Supercomputing resources at
NASA Ames Research Center.

\bibliographystyle{apj}

\begin{thebibliography}{42}
\expandafter\ifx\csname natexlab\endcsname\relax\def\natexlab#1{#1}\fi

\bibitem[{{Battaglia} {et~al.}(2005){Battaglia}, {Helmi}, {Morrison},
  {Harding}, {Olszewski}, {Mateo}, {Freeman}, {Norris}, \&
  {Shectman}}]{Battaglia05}
{Battaglia}, G., {et~al.} 2005, \mnras, 364, 433

\bibitem[{{Belokurov} {et~al.}(2007){Belokurov}, {Zucker}, {Evans}, {Kleyna},
  {Koposov}, {Hodgkin}, {Irwin}, {Gilmore}, {Wilkinson}, {Fellhauer},
  {Bramich}, {Hewett}, {Vidrih}, {De Jong}, {Smith}, {Rix}, {Bell}, {Wyse},
  {Newberg}, {Mayeur}, {Yanny}, {Rockosi}, {Gnedin}, {Schneider}, {Beers},
  {Barentine}, {Brewington}, {Brinkmann}, {Harvanek}, {Kleinman}, {Krzesinski},
  {Long}, {Nitta}, \& {Snedden}}]{Belokurov07}
{Belokurov}, V., {et~al.} 2007, \apj, 654, 897

\bibitem[{{Besla} {et~al.}(2007){Besla}, {Kallivayalil}, {Hernquist},
  {Robertson}, {Cox}, {van der Marel}, \& {Alcock}}]{Besla07}
{Besla}, G., {Kallivayalil}, N., {Hernquist}, L., {Robertson}, B., {Cox},
  T.~J., {van der Marel}, R.~P., \& {Alcock}, C. 2007, \apj, 668, 949

\bibitem[{{Besla} {et~al.}(2010){Besla}, {Kallivayalil}, {Hernquist}, {van der
  Marel}, {Cox}, \& {Kere{\v s}}}]{Besla10}
{Besla}, G., {Kallivayalil}, N., {Hernquist}, L., {van der Marel}, R.~P.,
  {Cox}, T.~J., \& {Kere{\v s}}, D. 2010, \apjl, 721, L97

\bibitem[{{Boylan-Kolchin} {et~al.}(2010){Boylan-Kolchin}, {Besla}, \&
  {Hernquist}}]{Boylan-Kolchin10}
{Boylan-Kolchin}, M., {Besla}, G., \& {Hernquist}, L. 2010, ArXiv e-prints

\bibitem[{{Busha} {et~al.}(2007){Busha}, {Evrard}, \& {Adams}}]{Busha07}
{Busha}, M.~T., {Evrard}, A.~E., \& {Adams}, F.~C. 2007, \apj, 665, 1

\bibitem[{{Busha} {et~al.}(2010){Busha}, {Wechsler}, \& {Behroozi}}]{Busha10}
{Busha}, M.~T., {Wechsler}, R.~H., \& {Behroozi}, P. 2010, {\it accepted to ApJ}, arXiv:1011.6373

\bibitem[{{Connors} {et~al.}(2006){Connors}, {Kawata}, \& {Gibson}}]{Connors06}
{Connors}, T.~W., {Kawata}, D., \& {Gibson}, B.~K. 2006, \mnras, 371, 108

\bibitem[{{Conroy} {et~al.}(2006){Conroy}, {Wechsler}, \&
  {Kravtsov}}]{Conroy06}
{Conroy}, C., {Wechsler}, R.~H., \& {Kravtsov}, A.~V. 2006, \apj, 647, 201

\bibitem[{{Dehnen} {et~al.}(2006){Dehnen}, {McLaughlin}, \&
  {Sachania}}]{Dehnen06}
{Dehnen}, W., {McLaughlin}, D.~E., \& {Sachania}, J. 2006, \mnras, 369, 1688

\bibitem[{{D'Onghia} \& {Lake}(2008)}]{DOnghia08}
{D'Onghia}, E., \& {Lake}, G. 2008, \apjl, 686, L61

\bibitem[{{Gardiner} {et~al.}(1994){Gardiner}, {Sawa}, \&
  {Fujimoto}}]{Gardiner94}
{Gardiner}, L.~T., {Sawa}, T., \& {Fujimoto}, M. 1994, \mnras, 266, 567

\bibitem[{{Gnedin} {et~al.}(2010){Gnedin}, {Brown}, {Geller}, \&
  {Kenyon}}]{Gnedin10}
{Gnedin}, O.~Y., {Brown}, W.~R., {Geller}, M.~J., \& {Kenyon}, S.~J. 2010,
  \apjl, 720, L108

\bibitem[{{Harris} \& {Zaritsky}(2006)}]{Harris06}
{Harris}, J., \& {Zaritsky}, D. 2006, \aj, 131, 2514

\bibitem[{{Kallivayalil} {et~al.}(2006{\natexlab{a}}){Kallivayalil}, {van der
  Marel}, \& {Alcock}}]{Kallivayalil06b}
{Kallivayalil}, N., {van der Marel}, R.~P., \& {Alcock}, C. 2006{\natexlab{a}},
  \apj, 652, 1213

\bibitem[{{Kallivayalil} {et~al.}(2006{\natexlab{b}}){Kallivayalil}, {van der
  Marel}, \& {Alcock}}]{Kallivayalil06}
---. 2006{\natexlab{b}}, \apj, 652, 1213

\bibitem[{{Kallivayalil} {et~al.}(2006{\natexlab{c}}){Kallivayalil}, {van der
  Marel}, {Alcock}, {Axelrod}, {Cook}, {Drake}, \& {Geha}}]{Kallivayalil06a}
{Kallivayalil}, N., {van der Marel}, R.~P., {Alcock}, C., {Axelrod}, T.,
  {Cook}, K.~H., {Drake}, A.~J., \& {Geha}, M. 2006{\natexlab{c}}, \apj, 638,
  772

\bibitem[{{Klypin} \& {Holtzman}(1997)}]{Klypin97}
{Klypin}, A., \& {Holtzman}, J. 1997, ArXiv Astrophysics e-prints

\bibitem[{{Klypin} {et~al.}(2010){Klypin}, {Trujillo-Gomez}, \&
  {Primack}}]{Klypin10}
{Klypin}, A., {Trujillo-Gomez}, S., \& {Primack}, J. 2010, ArXiv e-prints

\bibitem[{{Klypin} {et~al.}(2002){Klypin}, {Zhao}, \& {Somerville}}]{Klypin02}
{Klypin}, A., {Zhao}, H., \& {Somerville}, R.~S. 2002, \apj, 573, 597

\bibitem[{{Kravtsov} {et~al.}(2004){Kravtsov}, {Berlind}, {Wechsler}, {Klypin},
  {Gottl{\"o}ber}, {Allgood}, \& {Primack}}]{Kravtsov04}
{Kravtsov}, A.~V., {Berlind}, A.~A., {Wechsler}, R.~H., {Klypin}, A.~A.,
  {Gottl{\"o}ber}, S., {Allgood}, B., \& {Primack}, J.~R. 2004, \apj, 609, 35

\bibitem[{{Kravtsov} {et~al.}(1997){Kravtsov}, {Klypin}, \&
  {Khokhlov}}]{Kravtsov97}
{Kravtsov}, A.~V., {Klypin}, A.~A., \& {Khokhlov}, A.~M. 1997, \apjs, 111, 73

\bibitem[{{Kuhlen} {et~al.}(2010){Kuhlen}, {Weiner}, {Diemand}, {Madau},
  {Moore}, {Potter}, {Stadel}, \& {Zemp}}]{Kuhlen10}
{Kuhlen}, M., {Weiner}, N., {Diemand}, J., {Madau}, P., {Moore}, B., {Potter},
  D., {Stadel}, J., \& {Zemp}, M. 2010, JCAP, 2, 30

\bibitem[{{Lewis} \& {Bridle}(2002)}]{L+B02}
{Lewis}, A., \& {Bridle}, S. 2002, \prd, 66, 103511

\bibitem[{{Li} \& {White}(2008)}]{LiWhite08}
{Li}, Y., \& {White}, S.~D.~M. 2008, \mnras, 384, 1459

\bibitem[{{Lin} \& {Lynden-Bell}(1982) Lin \& Lynden-Bell}]{Lin82}
{Lin}, D.~N.~C. and {Lynden-Bell}, D. 1982, \mnras, 198, 707

\bibitem[{{Lisanti} {et~al.}(2010){Lisanti}, {Strigari}, {Wacker}, \&
  {Wechsler}}]{Lisanti10}
{Lisanti}, M., {Strigari}, L.~E., {Wacker}, J.~G., \& {Wechsler}, R.~H. 2010,
  ArXiv e-prints

\bibitem[{{Liu} {et~al.}(2010){Liu}, Gerke, Wechsler, Behroozi, \&
  Busha}]{Liu10}
{Liu}, L., Gerke, B., Wechsler, R., Behroozi, P., \& Busha, M. 2011, \apj, 733, 62 

\bibitem[{{MacKay} (2003)}]{MacKay}
{MacKay}, D.~J.~C.  2003, ``Information Theory, Inference and Learning Algorithms''. Cambridge: CUP

\bibitem[{{Murai} \& {Fujimoto}(1980)}]{Murai80}
{Murai}, T., \& {Fujimoto}, M. 1980, \pasj, 32, 581

\bibitem[{{Neal} (2001)}]{Neal01}
{Neal}, R.~M. 2001, Statistics and Computing, 11, 125

\bibitem[{{Piatek} {et~al.}(2008){Piatek}, {Pryor}, \& {Olszewski}}]{Piatek08}
{Piatek}, S., {Pryor}, C., \& {Olszewski}, E.~W. 2008, \aj, 135, 1024

\bibitem[{{Smith} {et~al.}(2007){Smith}, {Ruchti}, {Helmi}, {Wyse},
  {Fulbright}, {Freeman}, {Navarro}, {Seabroke}, {Steinmetz}, {Williams},
  {Bienaym{\'e}}, {Binney}, {Bland-Hawthorn}, {Dehnen}, {Gibson}, {Gilmore},
  {Grebel}, {Munari}, {Parker}, {Scholz}, {Siebert}, {Watson}, \&
  {Zwitter}}]{Smith07}
{Smith}, M.~C., {et~al.} 2007, \mnras, 379, 755

\bibitem[{{Stanimirovi{\'c}} {et~al.}(2004){Stanimirovi{\'c}},
  {Staveley-Smith}, \& {Jones}}]{Stanimirovic04}
{Stanimirovi{\'c}}, S., {Staveley-Smith}, L., \& {Jones}, P.~A. 2004, \apj,
  604, 176

\bibitem[{{Strigari} {et~al.}(2007){Strigari}, {Bullock}, {Kaplinghat},
  {Diemand}, {Kuhlen}, \& {Madau}}]{Strigari07d}
{Strigari}, L.~E., {Bullock}, J.~S., {Kaplinghat}, M., {Diemand}, J., {Kuhlen},
  M., \& {Madau}, P. 2007, \apj, 669, 676

\bibitem[{{Strigari} \& {Trotta}(2009)}]{StrigariTrotta}
{Strigari}, L.~E., \& {Trotta}, R. 2009, Journal of Cosmology and Particle
  Physics, 11, 19

\bibitem[{{Suyu} {et~al.}(2010){Suyu}, {Marshall}, {Auger}, {Hilbert},
  {Blandford}, {Koopmans}, {Fassnacht}, \& {Treu}}]{Suy10}
{Suyu}, S.~H., {Marshall}, P.~J., {Auger}, M.~W., {Hilbert}, S., {Blandford},
  R.~D., {Koopmans}, L.~V.~E., {Fassnacht}, C.~D., \& {Treu}, T. 2010, \apj,
  711, 201

\bibitem[{{Tinker} {et~al.}(2008){Tinker}, {Kravtsov}, {Klypin}, {Abazajian}, 
	{Warren}, {Yepes}, {Gottl{\"o}ber}, \& {Holz}}]{Tinker08}
{Tinker}, J., {Kravtsov}, A.~V., {Klypin}, A., {Abazajian}, K., {Warren}, M., {Yepes}, G., {Gottl{\"o}ber}, S., \& {Holz}, D.~E., \apj, 688, 709

\bibitem[{{Tollerud} {et~al.}(2008){Tollerud}, {Bullock}, {Strigari}, \&
  {Willman}}]{Tollerud08}
{Tollerud}, E.~J., {Bullock}, J.~S., {Strigari}, L.~E., \& {Willman}, B. 2008,
  \apj, 688, 277

\bibitem[{{Tollerud} {et~al.}(2011){Tollerud}, {Boylan-Kolchin}, {Barton}, {Bullock} \&
  {Trinh}}]{Tollerud11}
{Tollerud}, E.~J., {Boylan-Kolchin}, M., {Barton}, E.~S., {Bullock}, \& {Trinh}, C.~Q. 2011,
  arXiv:1103.1875

\bibitem[{{Trujillo-Gomez} {et~al.}(2010){Trujillo-Gomez}, {Klypin}, {Primack},
  \& {Romanowsky}}]{TrujilloGomez10}
{Trujillo-Gomez}, S., {Klypin}, A., {Primack}, J., \& {Romanowsky}, A.~J. 2010,
  ArXiv e-prints, arXiv:1005.1289


\bibitem[{{van den Bergh}(2000)}]{vandenBergh00}
{van den Bergh}, S. 2000, {The Galaxies of the Local Group}, ed. {van den
  Bergh, S.} (Cambridge)

\bibitem[{{van der Marel} {et~al.}(2002){van der Marel}, {Alves}, {Hardy}, \&
  {Suntzeff}}]{vanderMarel02}
{van der Marel}, R.~P., {Alves}, D.~R., {Hardy}, E., \& {Suntzeff}, N.~B. 2002,
  \aj, 124, 2639

\bibitem[{{Vogelsberger} {et~al.}(2009){Vogelsberger}, {Helmi}, {Springel},
  {White}, {Wang}, {Frenk}, {Jenkins}, {Ludlow}, \& {Navarro}}]{Vogelsberger09}
{Vogelsberger}, M., {et~al.} 2009, \mnras, 395, 797

\bibitem[{{Walsh} {et~al.}(2009){Walsh}, {Willman}, \& {Jerjen}}]{Walsh09}
{Walsh}, S.~M., {Willman}, B., \& {Jerjen}, H. 2009, \aj, 137, 450

\bibitem[{{Watkins} {et~al.}(2010){Watkins}, {Evans}, \& {An}}]{Watkins10}
{Watkins}, L.~L., {Evans}, N.~W., \& {An}, J.~H. 2010, \mnras, 406, 264

\bibitem[{{Wechsler} {et~al.}(2002){Wechsler}, {Bullock}, {Primack},
  {Kravtsov}, \& {Dekel}}]{Wechsler02}
{Wechsler}, R.~H., {Bullock}, J.~S., {Primack}, J.~R., {Kravtsov}, A.~V., \&
  {Dekel}, A. 2002, \apj, 568, 52

\bibitem[{{Xue} {et~al.}(2008){Xue}, {Rix}, {Zhao}, {Re Fiorentin}, {Naab},
  {Steinmetz}, {van den Bosch}, {Beers}, {Lee}, {Bell}, {Rockosi}, {Yanny},
  {Newberg}, {Wilhelm}, {Kang}, {Smith}, \& {Schneider}}]{Xue08}
{Xue}, X.~X., {et~al.} 2008, \apj, 684, 1143

\end{thebibliography}

\end{document}